# Synthesis and Structural Analysis of Multilayered Graphene via Microwave Atmospheric Pressure Plasma

Waqas Iqbal[1], Najeeb-ur-Rehman[1*], Niaz Wali[1,2]

[1] Plasma Physics Laboratory, Department of Physics, COMSATs University, Islamabad, Pakistan

[2] Institute for Fusion Theory and Simulation, School of Physics, Zhejiang University, Hangzhou, 310058, China

Email of the corresponding author: najeeb-ur-rehman@comsats.edu.pk

**ABSTRACT−** This study reports the successful synthesis of multilayered graphene sheets via microwave atmospheric pressure plasma. This innovative approach streamlines and expedites graphene production and other carbon nanostructures, eliminating the need for catalysts, solvents, or complex processing conditions. Ethanol is directly injected into a microwave-generated argon plasma plume, leading to the formation of graphene. Raman spectroscopy revealed characteristic peaks (2D, G, and D bands) confirming graphene composition, with defects indicated by the D band. X-ray diffraction analysis supported these findings, indicating a broad peak at 25º corresponding to the (002) plane, affirming a multi-layered graphene structure. Scanning electron microscopy exhibited crumpled, randomly oriented graphene sheets, albeit with uneven structures suggesting impurity incorporation. The presence of defects was quantified through the intensity ratio of the D to G band ($I_D/I_G$) in Raman spectroscopy, revealing a value of 0.80, signifying the presence of defects in the synthesized graphene. The 2D to G band intensity ratio ($I_{2D}/I_G$) suggested the existence of 7-10 graphene layers, highlighting the need for further optimization for enhanced graphene quality and purity.

**Keywords:** Graphene, Synthesis, Microwave plasma, Raman spectroscopy, Morphological analysis

## I.  INTRODUCTION

The two-dimensional crystalline substance graphene, an allotrope of carbon characterized by a one-atom-thick, flat structure, has received intensifying attention since it was discovered in 2004 [1]. This surge of interest is fueled by both the tangible potential for practical applications and the extraordinary electronic characteristics exhibited by this material. Research on graphene and related structures like sheets, nanoribbons, and carbon nanotubes has made considerable progress in recent years, driven by their outstanding properties and wide-ranging potential in applications like gas sensors, super-capacitors, batteries, electrodes, solar cells, transistors, thermoelectricity, memory storage devices, self-healing materials, and spacecraft technology [2-13].

Developing an efficient processing technique is crucial for uncovering the manifold attributes of graphene and unlocking its full potential in various applications. Several approaches have been employed to synthesize graphene, encompassing micromechanical cleavage of bulk graphite, chemical exfoliation, chemical vapor deposition (CVD), solvothermal techniques, thermal decomposition of silicon carbide, plasma-enhanced chemical vapor deposition (PECVD), and gas-phase synthesis via microwave plasma [14-20]. The CVD enables large-scale graphene synthesis, often yielding low-quality graphene. It necessitates optimal energy levels and balanced carbon particle influx for high-quality graphene [21]. While chemical exfoliation and solvothermal methods are both effective for large-scale graphene production, chemical processes may compromise the quality of the resultant graphene [22, 23]. On the other hand, gas phase synthesis utilizing microwave plasma is not a chemical synthesis method, and mass production is achievable. However, the average yield is around 1.2 percent, which is extremely low [24].

The synthesis of graphene using plasma is a recently developed approach with little existing literature [25-31]. The ability to manufacture self-standing graphene is the most significant benefit of plasma-based synthesis over other known methods. The reactive chemical plasma atmosphere is ideal for developing different synthetic routes that result in consistent structures [8]. The Surfaguide microwave (MW) system is a single-step approach for producing pure graphene sheets of excellent quality. This method provides ideal conditions for nucleation and growth for synthesizing graphene, carbon nanotubes (CNTs), and graphene quantum dots (GQD). The breakdown of ethanol can be used to make graphene via a bottom-up approach. In 2013, E. Tatarova et al. successfully synthesized graphene multi-layers using the Surfaguide MW system [25].

In this work, we reported the synthesis of graphene via Surfaguide MW systemin which ethanol is introduced via a syringe to the argon-grazed plasma plume at atmospheric pressure. We have systematically optimized the Surfaguide MW system for graphene synthesis, closely monitoring discharge parameters (such as microwave power, gas flow rate, and gas concentration). The synthesized graphene sheets were characterized using Raman spectroscopy, X-ray diffraction (XRD), and scanning electron microscopy (SEM). The paper is organized as follows: The experimental procedure is described in Sec. II. Experimental results are discussed in Sec. III. Section IV is the conclusion.

## II.  EXPERIMENTAL PROCEDURE

The experiment is carried out in microwave based Surfaguide configuration available at

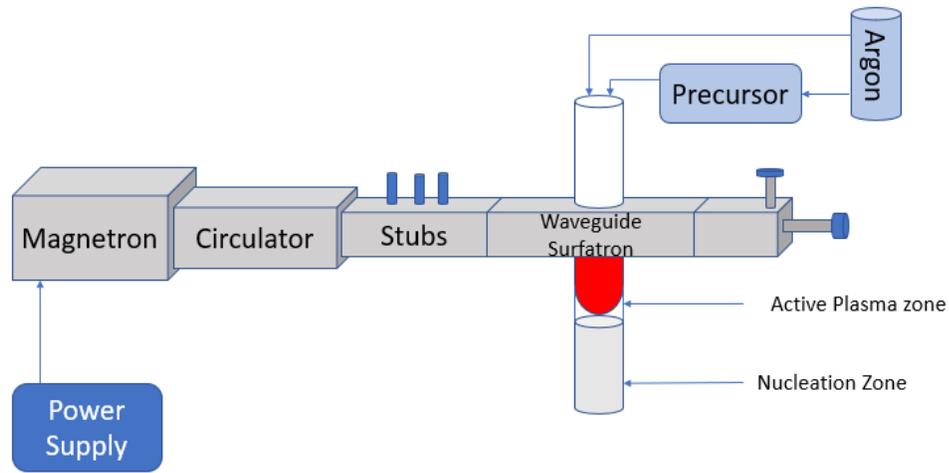

*Fig. 1.  Schematic diagram of Surfaguide microwave plasma system used to synthesize*

Plasma Research Laboratory. An atmospheric pressure argon plasma is generated in a quartz tube-made reactor (20 mm thick) connected with a cooling unit to run the overall system smoothly, as depicted in Figure 1. The setup comprised a magnetron powered by a 2.45GHz and 3kW

Microwave source to generate microwaves, which, along with wave conditioning components, were guided toward the reactor tube via the WR 340 Tapered waveguide system. Precise control over these parameters was instrumental in achieving the optimal plasma conditions for successful graphene synthesis. The argon gas flow rate varied from 1000 to 1700 SCCM, and the MW power was adjusted between 900W and 2 kW. This experimental design allowed for real-time adjustments, enabling researchers to fine-tune the process for optimal results. The system was optimized at 1700 SSCM argon flow rate and 900 W microwave power source.

Ethanol precursor was consistently delivered at a rate of 3.6 mL/h for a given argon flow rate, ensuring a steady and reliable source for the graphene synthesis process. Under these conditions, graphene synthesis was monitored concerning discharge parameters such as MW power, gas flow rate, and precursor fraction. Elevated temperatures in the discharge zone facilitated ethanol decomposition into reactive hydrocarbon fragments, which subsequently nucleated and grew in the plasma afterglow, forming graphene in aromatic carbon rings. The final product was collected on a nylon filter and glass plate, guaranteeing a high graphene sheet yield and minimizing potential loss. The discharge and Surfaguide MW system were cooled to maintain stability using water and air. This meticulous combination of precise monitoring and cooling measures ensured successful experiments and contributed to the equipment's longevity, allowing for a thorough study of graphene synthesis. Finally, the synthesized graphene sheets were characterized with Raman spectroscopy, X-ray diffraction (XRD), and scanning electron microscopy (SEM).

### III. EXPERIMENTAL RESULTS

Surfaguide microwave plasma was used to synthesize graphene samples at atmospheric pressure. Graphene was produced at various microwave forward powers and argon flow rates while the ethanol concentration remained constant throughout the experiment. Different graphene samples are shown in figures 2. Graphene samples were acquired and subsequently deposited on a nylon filter and a glass plate. Following deposition, the samples underwent characterization using techniques such as scanning electron microscopy (SEM), Raman spectroscopy, and X-ray diffraction (XRD) to confirm the presence and quality of graphene. The collected graphene samples can be employed for specific applications or further studies, considering graphene's unique properties that make it suitable for various purposes in fields such as electronics and materials science.

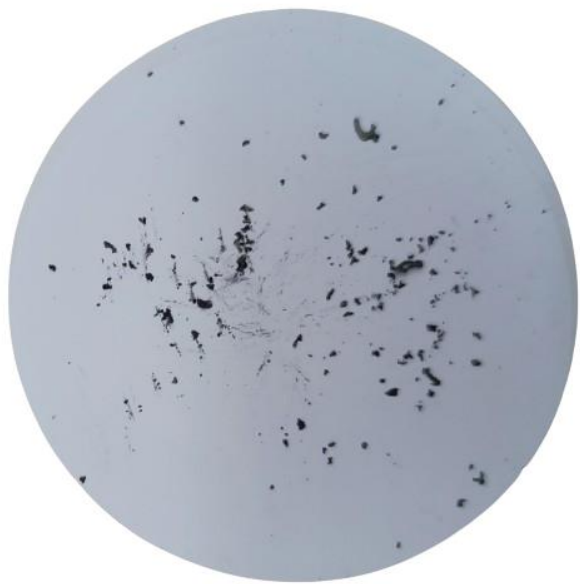

*Fig. 2. The collected graphene samples on the nylon filter.*

i. **RAMAN SPECTROSCOPY ANALYSIS**

Raman spectroscopy of the prepared graphene sample through microwave atmospheric pressure plasma is performed using the Raman spectrometer model Dong Woo with an excitation wavelength of 514 nm utilizing an argon ion laser as an excitation source. During the measurement, lens magnification is 100X, exposure time is one second, and number of accumulations is 100. The experiment is performed at room temperature. The Raman spectrum of the synthesized graphene sample on the nylon filter through microwave argon plasma is shown in Figure 3. Three distinct peaks arise at different Raman positions in the Raman spectra of the synthesized graphene sample.

The three peaks, corresponding to the 2D, G, and D bands, occurred at 2746 cm$^{-1}$, 1605 cm$^{-1}$, and 1374 cm$^{-1}$, respectively. This form of Raman spectrum is commonly connected with graphene, indicating that the material obtained is graphene. The appearance of the D band implies that the graphene sheet has defects. The D band appears due to imperfections in the graphene induced by dangling bonds in the edges and broken layers in the structure. The G band indicates the carbon atoms' in-plane vibrational modes, originating from the covalent bond stretching of the hexagonal aromatic rings in the graphene sheet. The 2D band, solid evidence for graphene identification and specifies the number of graphene layers, is the most vital band in the Raman spectrum.

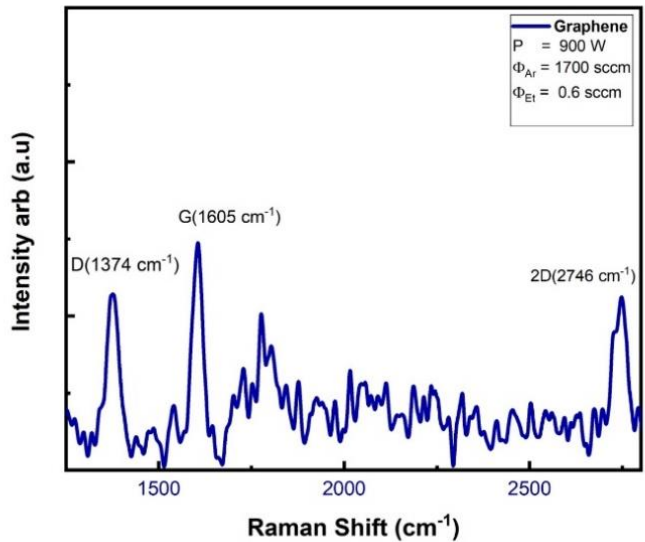

*Fig. 3. Raman spectrum of the synthesized graphene sheet obtained on nylon filter using microwave argon plasma at atmospheric pressure.*

Due to the defects in the synthesized graphene sample, there is a disorder in the graphene structure. The intensity ratios of the D to G band (ID/IG) estimate disease in graphene structure. For the synthesized graphene sample, the intensity ratio of the D to G band is 0.80, which indicates that defects are present in synthesized graphene. The number of graphene layers is evaluated from the intensity ratio of the 2D to G band (I2D/IG). The intensity ratio of the 2D to G band shows that the synthesized graphene sample has multi-layers (7-10) of graphene due to the adjacent layers of graphene reabsorbing few-layer graphene scattered photons. As a result, a small number of scattered photons exist. Therefore, the intensity of the 2D band is low. The power of the 2D peak drops as the number of graphene layers increases, and the peak becomes broader. Data from the Raman spectroscopy of the synthesized graphene sample is shown in Table 1.

| Sr. No | Sample | Substrate | D-Band (cm$^{-1}$) | G-Band (cm$^{-1}$) | 2D-Band (cm$^{-1}$) | $\frac{I_D}{I_G}$ | $2\frac{I_{2D}}{I_G}$ |
|---|---|---|---|---|---|---|---|
| 1 | 1 | Nylon Filter | 1374 | 1605 | 2746 | 0.80 | 0.79 |

**Table 1:** Raman data of the synthesized graphene sheet using microwave atmospheric pressure plasma.

### ii. X-RAY DIFFRACTION (XRD) ANALYSIS

X-ray diffraction was employed to investigate the crystal structure of graphene. Figure 3 shows the XRD pattern of graphene prepared by ethanol decomposition with microwave argon plasma at atmospheric pressure using a surface wave reactor. The X-ray diffraction pattern of graphene was obtained with Panalytical company model X'Pert PRO X-ray diffractometer operating at 30 mA operating current and voltage 40 V with a 1.54 Å (Cu-kα) wavelength radiation as an X-ray source.

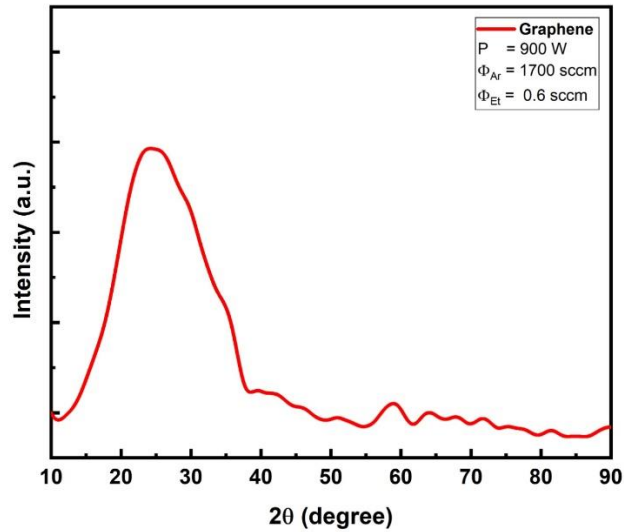

*Fig. 4. XRD diffraction pattern of the synthesized graphene sheet obtained on glass plate using microwave argon plasma at atmospheric pressure with 1kW microwave power and 1700 SCCM argon gas flow.*

The device operates at a scan rate of 120 seconds. Data was obtained at room temperature with a 2θ range between 10 to 90.

The XRD pattern in Figure 4 shows a broad peak at 2θ value of 25º for graphene, which is the same peak as in graphite compared with the JCPDS card (75-1621) and corresponds to a (002) plane. From the observed peak position, the inter-planner distance can be calculated, and the result found was 0.34 nm, the same space as in the hexagonal graphite structure. This confirms the synthesis of graphene and indicates that the graphene is a multi-layer graphene, which matches well with the Raman results. The observed peak is not sharp but comprehensive, and its intensity is low, indicating that the created graphene sheet has a structure between crystalline and amorphous. The observed peak broadening could be caused by a mixture of graphene sheets with different layers, indicating that the crystal phase (002) is not well arranged. Table 2 shows the value of the 2θ, plane, and the inter-planner spacing (d) obtained from the XRD pattern of the graphene by using the Expert high score software.

| Sr. No | Sample | Substrate | 2θ (Degree) | plane | d-spacing (nm) |
| --- | --- | --- | --- | --- | --- |
| 1 | 1 | Glass Plate | 25 | (002) | 0.34 |

**Table 2:** XRD data of the synthesized graphene sheet using microwave plasma at atmospheric pressure.

### iii. SCANNING ELECTRON MICROSCOPY (SEM) ANALYSIS

Scanning electron microscopy was used to examine the morphology of the synthesized graphene. The micrograph was taken at a magnification of X50,000 operated at 5 kV. The SEM micrograph of the synthesized graphene sample is shown in Figure 5.

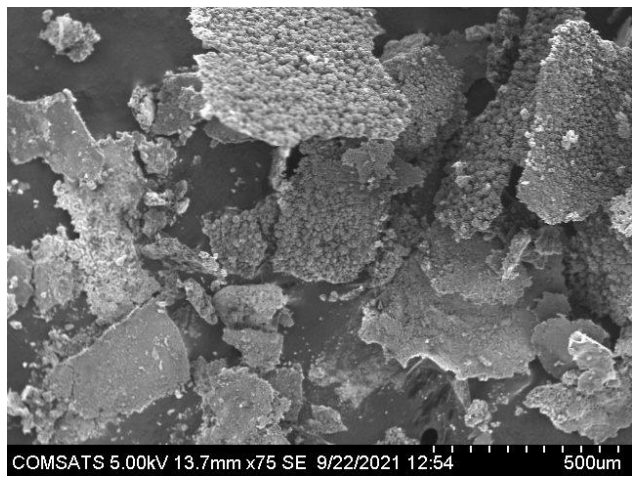

*Fig. 5. SEM micrograph of the graphene sample synthesized at 2 kW microwave power and 1700 SCCM argon flow.*

The morphology in the SEM micrograph suggests that multilayered graphene sheets are successfully produced. The sheets of graphene are crumpled and orientated randomly in the image below. Other uneven structures are also seen, implying that impurities have been incorporated into the graphene structure during the preparation. The contaminants may be due to other carbon structures and oxygen. As ethanol is a precursor for synthesizing graphene, the present impurities are expected as graphene synthesis is performed at atmospheric pressure. There are chances that other impurities are incorporated, possibly due to the incomplete dissociation of ethanol.

### IV. CONCLUSION

The successful synthesis of multilayered graphene sheets using microwave atmospheric pressure plasma has been confirmed through collective analysis of experimental results obtained via Raman spectroscopy, X-ray diffraction (XRD), and scanning electron microscopy (SEM). The Raman spectra revealed characteristic peaks corresponding to the 2D, G, and D bands, affirming the graphene nature of the synthesized material. The presence of the D band indicated the existence of defects, while the low intensity of the 2D band suggested a multi-layered graphene structure (7-10 layers). The XRD pattern further supported these findings, which showed a broad peak at a 2θ value of 25º corresponding to the (002) graphene plane, indicating a multi-layered structure with an inter-planar distance of 0.34 nm. SEM analysis provided visual evidence of crumpled and randomly oriented multilayered graphene sheets. However, the presence of uneven structures implied the incorporation of impurities, potentially originating from other carbon structures and oxygen, during the synthesis process. These impurities were attributed to using ethanol as a precursor for graphene synthesis at atmospheric pressure, resulting in incomplete dissociation and the incorporation of contaminants. The observed defects and impurities underscore the need to optimize further the graphene synthesis process, alternative precursors, and post-synthesis treatments for enhanced graphene quality and purity.

## ACKNOWLEDGMENT

The authors gratefully acknowledge the valuable financial support provided by the Higher Education Commission (HEC), Islamabad, Pakistan, for this work under Grant No. TDF-137.

## DECLARATIONS

**Conflict of interest** The authors do not have any relevant financial or non-financial interests to declare.

**DATA AVAILABILITY STATEMENT** This manuscript has no associated data. The data that support the findings of this study are available from the corresponding author upon reasonable request.